\documentclass[twocolumn,secnumarabic,amsmath, amssymb, nobibnotes, aps, prb]{revtex4-1}
\usepackage{listings}

\usepackage{subfigure}% Include figure files
\usepackage{graphicx}% Include figure files
\graphicspath{{./}{./images/}{./images_all/}{./images_all/spin_pumping_lambda/}}

\usepackage{dcolumn}% Align table columns on decimal point
\usepackage{bm}% bold math
\begin{document}

\title{Estimating spin diffusion length and spin Hall angle from spin pumping-induced inverse spin Hall voltages}% Force line breaks with \\

\author{Kuntal Roy}
\email{kuntalroy@gmail.com.}
%\noaffiliation
\affiliation{School of Electrical and Computer Engineering, Purdue University, West Lafayette, Indiana 47907, USA}

%\date{\today}% It is always \today, today,
             %  but any date may be explicitly specified
             %  but any date may be explicitly specified

\begin{abstract}
There exists considerable confusion in estimating the spin diffusion length of materials with high spin-orbit coupling from spin pumping experiments. For designing functional devices, it is important to determine the spin diffusion length with sufficient accuracy from experimental results. An inaccurate estimation of spin diffusion length also affects the estimation of other parameters (e.g., spin mixing conductance, spin Hall angle) concomitantly. The spin diffusion length for platinum (Pt) has been reported in literature in a wide range of 0.5 -- 14 nm, and particularly it is a constant value independent of Pt's thickness. Here, the key reasonings behind such wide range of reported values of spin diffusion length  have been identified comprehensively. Particularly, it is shown here that a thickness-dependent conductivity and spin diffusion length is necessary to simultaneously match the experimental results of effective spin mixing conductance and inverse spin Hall voltage due to spin pumping. Such thickness-dependent spin diffusion length is tantamount to Elliott-Yafet spin relaxation mechanism, which bodes well for transitional metals. This conclusion is not altered even when there is significant interfacial spin memory loss. Furthermore, the variations in the estimated parameters are also studied, which is important for technological applications.
\end{abstract}

%\pacs{}% PACS, the Physics and Astronomy
                             % Classification Scheme.
%\keywords{Spin pumping, spin diffusion length, spin Hall angle, Elliott-Yafet spin relaxation}
% Use showkeys class option if keyword
                              %display desired

\maketitle

\section{Introduction}

In spin pumping~\cite{RefWorks:1311,RefWorks:1305,*RefWorks:1306,*RefWorks:1304,*RefWorks:1307,*RefWorks:1308,*RefWorks:1309,*RefWorks:1310,RefWorks:881,*RefWorks:1041,RefWorks:876,RefWorks:882,RefWorks:989} mechanism, unlike charge pumping~\cite{RefWorks:1034,*RefWorks:1324}, a precessing magnet sustained by an externally applied alternating magnetic field~\cite{RefWorks:1168} emits \emph{pure} spins into surrounding conductors. According to Onsager's reciprocity~\cite{RefWorks:1292,*RefWorks:1293}, spin pumping is the reciprocal phenomenon~\cite{RefWorks:1295} of spin momentum transfer~\cite{RefWorks:8,*RefWorks:155,*RefWorks:7,*RefWorks:196}. Theoretical constructs~\cite{RefWorks:881,*RefWorks:1041,RefWorks:876} well support the experimental results on spin pumping~\cite{RefWorks:1305,*RefWorks:1306,*RefWorks:1304,*RefWorks:1307,RefWorks:1309,*RefWorks:1308,RefWorks:1310}. If the adjacent normal-metal possess high spin-orbit coupling~\cite{RefWorks:810,RefWorks:1352} (e.g., platinum~\cite{RefWorks:757}, tantalum~\cite{RefWorks:608}, tungsten~\cite{RefWorks:758}, CuIr~\cite{RefWorks:1341}, CuBi~\cite{RefWorks:755}, CuPb~\cite{RefWorks:1342}, AuW~\cite{RefWorks:1343}), a considerable amount of dc charge voltage can be generated allowing the detection of spin current via inverse spin Hall effect (ISHE)~\cite{RefWorks:1101,RefWorks:1325,RefWorks:1278,RefWorks:769,RefWorks:757,RefWorks:1011,RefWorks:1008}. Therefore, the spin pumping mechanism gives us an alternative methodology to understand and estimate the relevant parameters in the system. Such understandings can benefit the device design using SHE~\cite{roy14_3}, which has potential for building future spintronic devices, alongwith other promising emerging devices~\cite{roy16_spin}.

From the spin pumping experiments there are three parameters to quantify: spin mixing conductance ($g^{\uparrow \downarrow}$) at the ferromagnet-normal metal (FM-NM) interface~\cite{RefWorks:1039,*RefWorks:1040,RefWorks:1018}, spin diffusion length ($\lambda$), and spin Hall angle ($\theta_{SH}$) of the SHE layer acting as the NM layer. There are also interface resistance and spin flip parameter to identify when there is significant interfacial spin memory loss~\cite{RefWorks:1319,RefWorks:367,RefWorks:1338,RefWorks:1340}. Experimentally, we get two quantities: effective spin mixing conductance ($g_{eff}^{\uparrow \downarrow}$) of the whole structure from damping enhancement and the induced inverse spin Hall voltage ($V_{ISHE}$) due to inverse spin Hall effect. There exists controversy~\cite{RefWorks:1285,RefWorks:1286,RefWorks:983,RefWorks:1111,RefWorks:984,RefWorks:985,RefWorks:1287,RefWorks:1102,RefWorks:1020,RefWorks:1021,RefWorks:1303,RefWorks:1019,RefWorks:818,RefWorks:885,RefWorks:1114,RefWorks:1001,RefWorks:982,RefWorks:1014,RefWorks:1300,RefWorks:812,RefWorks:998,RefWorks:1022,RefWorks:1302,RefWorks:986,RefWorks:814,*RefWorks:813,RefWorks:1013,RefWorks:811,RefWorks:1288,RefWorks:980} in determining $\lambda$, which is reported in wide range 0.5 -- 14 nm. Such wide range creates a massive issue in designing and predicting device functionality.  We study the underlying theoretical constructs and address such open issue. In particular, experimental results for permalloy (Py)/platinum (Pt) bilayers in Ref.~\citenum{RefWorks:986} show that $g_{eff}^{\uparrow \downarrow}$ and a quantity dependent on $V_{ISHE}$ saturate at different thicknesses of the SHE layer ($\lambda$ $\sim$ 1.5 nm and 8.3 nm, respectively). However, Ref.~\citenum{RefWorks:983} estimated a considerably different $\lambda$ of 1.2 nm from experimental results of $V_{ISHE}$, but the $\lambda$ is fixed independent of the thickness of the Pt layer, according to the usual perception~\cite{RefWorks:1287}. Also, Ref.~\citenum{RefWorks:998} determines a $\lambda$ of 7.3 nm from $V_{ISHE}$ and Ref.~\citenum{RefWorks:980} determines a $\lambda$ of 0.8 nm from $g_{eff}^{\uparrow \downarrow}$ recently for Py/Pt bilayers. 

Here we analyze the key issues behind such major disagreements of  the estimated values of the $\lambda$ and its dependence of the SHE layer thickness. In room-temperature measurements, the issues are as follows: (1) The conductivities of the different samples used in different experiments are different. A higher conductivity $\sigma$ would lead to a higher $\lambda$ concomitantly ($\lambda \propto \sigma$), due to Elliott-Yafet spin relaxation mechanism~\cite{RefWorks:1297,*RefWorks:1298}, which is relevant for transition metals~\cite{RefWorks:1283,RefWorks:1013,RefWorks:1319,roy17_2,*roy_aps_2015}. (2) With Elliott-Yafet spin relaxation mechanism, $\lambda$ is dependent on thickness since conductivity varies with thickness of the sample (there is interface contribution as well)~\cite{RefWorks:885}. The Dyakonov-Perel spin relaxation mechanism~\cite{RefWorks:1296} corresponds to a constant $\lambda$, which is usually used but it is not the relevant spin relaxation mechanism for transitional metals. (3) It needs to use a correct (must be positive) and relevant bare $g^{\uparrow \downarrow}$. The $g^{\uparrow \downarrow}$ is not achievable directly from experiments, rather what is achieved is an effective conductance of the whole bilayer $g_{eff}^{\uparrow \downarrow}$, which is thickness-dependent. Also, the so-called Schep correction~\cite{RefWorks:1078,*RefWorks:1077,RefWorks:876,RefWorks:1281} needs to be performed on the the bare $g^{\uparrow \downarrow}$, which makes a large difference. It is shown here that the controversy acclaimed in the Ref.~\citenum{RefWorks:986} that two different constant values of spin diffusion lengths are required to explain the experimental results of $g_{eff}^{\uparrow \downarrow}$ and a quantity dependent on $V_{ISHE}$ can be solved by considering a thickness-dependent $\lambda$~\cite{roy17_2,*roy_aps_2015}.

Since the variations in the estimated parameters pose limitations in designing technological applications, the variations in spin diffusion length and spin Hall angle are studied with respect to the variations in the interfacial spin mixing conductance and spin memory loss, which may be different due to fabrication process and from sample to sample. It is found that the estimated parameters are quite sensitive on the interfacial conductances.

\begin{figure}
\centering
\includegraphics{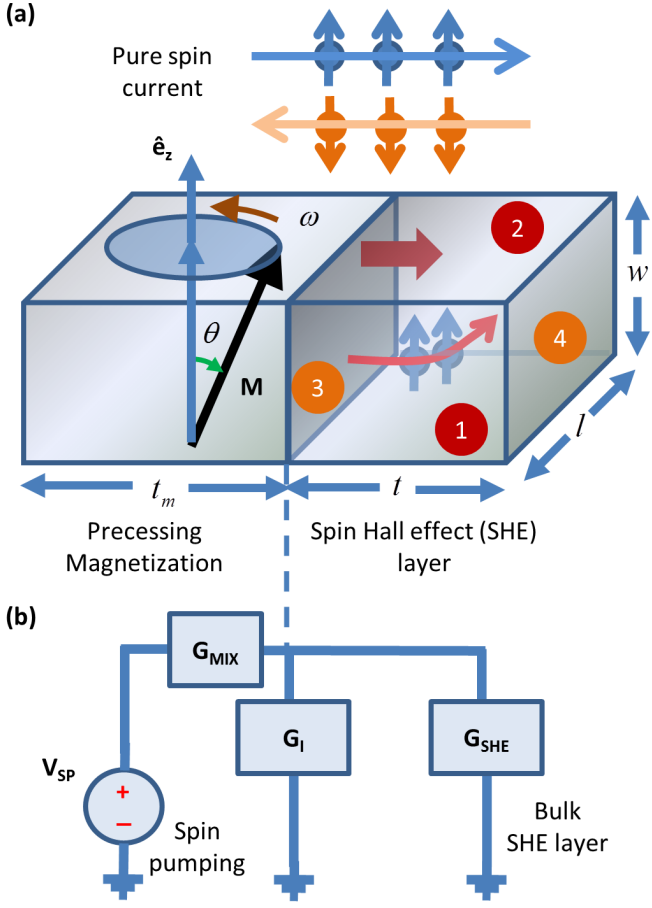}
\caption{\label{fig:spin_pumping_circuit} (a) A precessing magnetization in a magnetic layer is pumping pure spin current into the adjacent normal metal (NM). If the normal metal has a high spin orbit coupling, a considerable amount of charge current can be produced due to inverse spin Hall effect (ISHE). Spin potentials are developed at the surfaces marked by 3 and 4, while charge potentials are developed at the surfaces marked by 1 and 2. (b) In an equivalent spin circuit diagram, the voltage source $V_{SP}$ acts as a spin battery, $G_{MIX}$ is the interfacial spin mixing conductance between the magnetic layer and the SHE layer, $G_{I}$ is the spin conductance representing the spin memory loss, and $G_{SHE}$ is the spin conductance of the SHE layer, altered by the spin accumulation in the SHE layer in the presence of spin memory loss.}
\end{figure}

\begin{figure}[t]
\centering
\includegraphics{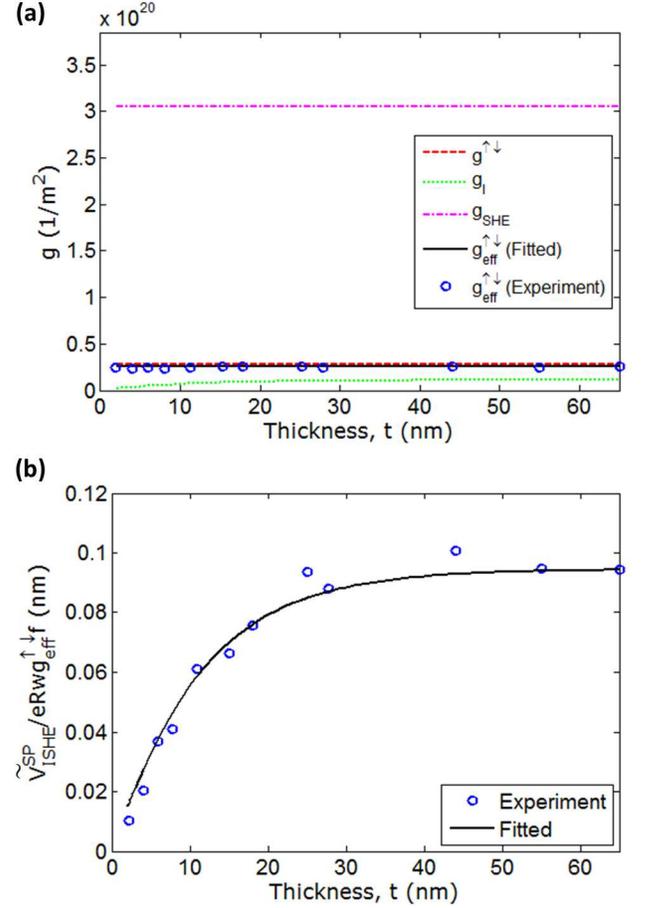}
\caption{\label{fig:match_experiments} Fitting the thickness $t$ (2--65 nm) dependence of (a) effective spin-mixing conductance $g_{eff}^{\uparrow \downarrow}$ [Equation~\eqref{eq:g_mix_eff}], and (b) a metric $\tilde{V}^{SP}_{ISHE}/e R w g_{eff}^{\uparrow \downarrow}f = \theta_{SH} \lambda \, tanh(t/\lambda)$ [Equation~\eqref{eq:V_tilde_SP_ISHE}]. Experimental data points are taken from the Ref.~\citenum{RefWorks:986}, precisely for f = 9 GHz. In part (a), the $g^{\uparrow \downarrow}$, $g_I$, and $g_{SHE}$ are plotted too.}
\end{figure}

\begin{figure}
\centering
\includegraphics{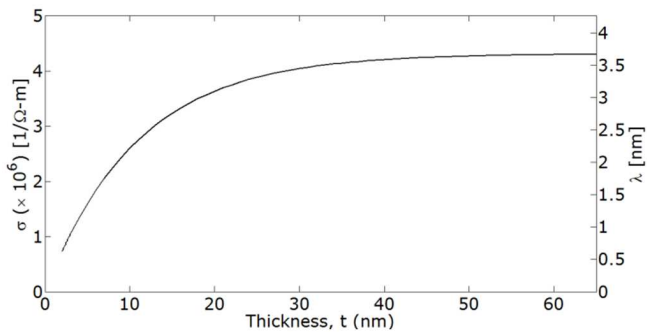}
\caption{\label{fig:lambda_sigma} Thickness $t$ (2--65 nm) dependence of conductivity and spin diffusion length $\lambda$. Due to Elliot-Yafet spin relaxation mechanism, $\lambda \propto \sigma$, and $\sigma$ depends on the thickness $t$. The saturated value $\lambda_{max}$ ($\sigma_{max}$) = 3.67 nm (4.3e6 1/$\Omega$-m) and the lowest value of $\lambda$ ($\sigma$) = 0.62 nm (0.73e6 1/$\Omega$-m) at $t=2$ nm.}
\end{figure}

\begin{figure*}
\centering
\includegraphics{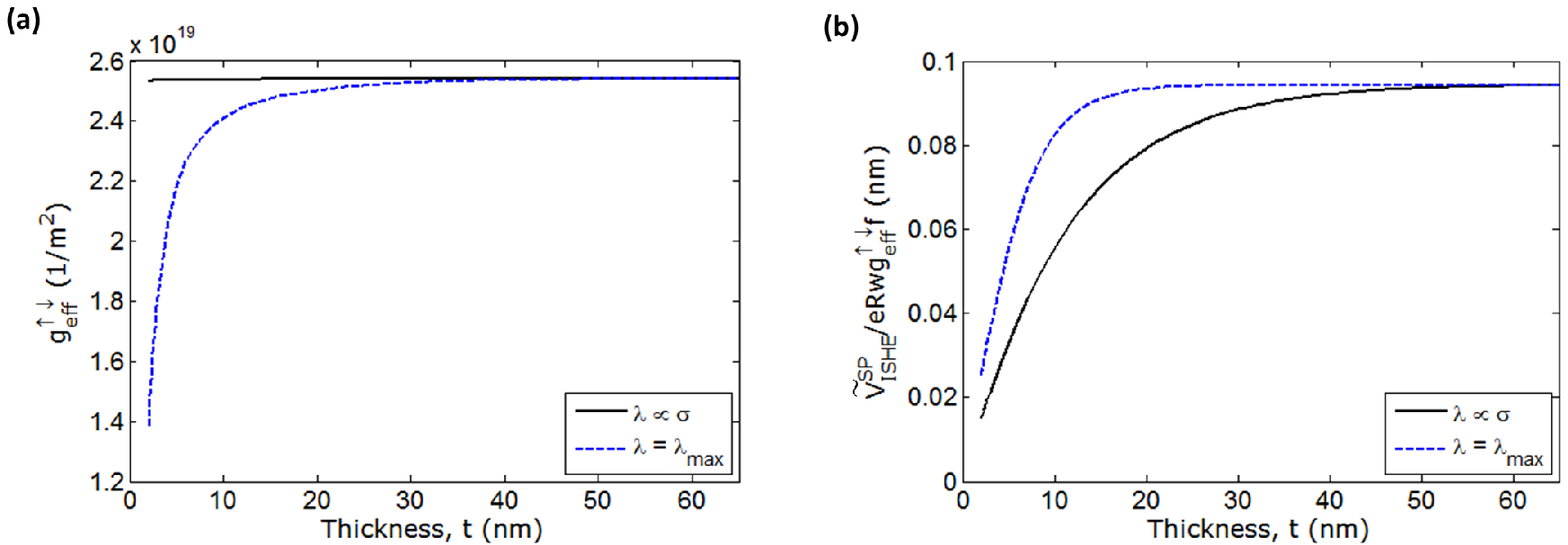}
\caption{\label{fig:lambda_constant_compare} Thickness $t$ (2--65 nm) dependence of (a) effective spin-mixing conductance $g_{eff}^{\uparrow \downarrow}$ [Equation~\eqref{eq:g_mix_eff}], and (b) a metric $\tilde{V}^{SP}_{ISHE}/e R w g_{eff}^{\uparrow \downarrow}f = \theta_{SH} \lambda \, tanh(t/\lambda)$ [Equation~\eqref{eq:V_tilde_SP_ISHE}] for two cases: $\lambda \propto \sigma$ (Elliott-Yafet spin relaxation mechanism) and $\lambda = \lambda_{max} = 3.67$ nm (Dyakonov-Perel spin relaxation mechanism). 
}
\end{figure*}

\section{Model}
\label{sc:model}

Figure~\ref{fig:spin_pumping_circuit}(a) shows a schematic diagram for spin pumping by a precessing magnetization into a SHE layer having a length $l$, width $w$, and thickness $t$. The thickness of the magnet is $t_m$. The corresponding spin circuit representation~\cite{roy17_2,*roy_aps_2015} containing voltage source and conductances~\cite{RefWorks:1284,*RefWorks:1289,*RefWorks:1290} is shown in the Fig.~\ref{fig:spin_pumping_circuit}(b). The voltage source $V_{SP}$ acts as a spin battery~\cite{RefWorks:878,*RefWorks:1038}, $G_{MIX}$ is the interfacial bare spin mixing conductance between the magnetic layer and the SHE layer~\cite{RefWorks:1039,*RefWorks:1040,RefWorks:1018}, $G_I$ represents the spin conductance due to spin memory loss with parameter $\delta$ representing the spin flip probability $1-e^{-\delta}$ at the interface, and $G_{SHE}$ is the spin conductance of the SHE layer, altered by the spin accumulation in the SHE layer in the presence of spin memory loss~\cite{RefWorks:1319,RefWorks:367,RefWorks:1338,RefWorks:1340}. The conductances $G_{MIX}$, $G_{I}$, and $G_{SHE}$ per unit area are defined as $G_{MIX}/lw = (2e^2/h)\, g^{\uparrow \downarrow}$, $G_{I}/lw = (\delta/R^*)\,sinh(\delta) = (2e^2/h)\, g_{I}$, and $G_{SHE}/lw = (\sigma/\lambda)\, cosh(\delta) \, tanh(t/\lambda) = (2e^2/h)\, g_{SHE}$, where $R^*$ is an effective interface resistance depending on the interface spin polarization~\cite{RefWorks:1319,RefWorks:367,RefWorks:1338,RefWorks:1340}, $\sigma$ and $\lambda$ are the conductivity and spin diffusion length of the SHE layer, respectively, and the conductances $g^{\uparrow \downarrow}$, $g_I$, and $g_{SHE}$ are in the units of $m^{-2}$. Note that $R^*$ depends on $\sigma$~\cite{RefWorks:1319}. The conductance $g_{SHE}$ takes care of the backflow of the accumulated spins in the SHE layer~\cite{RefWorks:881,*RefWorks:1041,RefWorks:876}.

The interfacial spin mixing conductance for FM-NM bilayers can be determined from first principles~\cite{RefWorks:1046} as $\tilde{g}^{\uparrow \downarrow}=g_{sh}-\sum{r_{mn}^{\uparrow}r_{mn}^{\downarrow *}}$ (where $g_{sh}$ is the so-called Sharvin conductance, i.e., the number of transport channels per unit area for one spin, and $r_{mn}^s$ is the probability amplitude of reflection from channel $n$ to channel $m$ with same spin $s$)~\cite{RefWorks:1039,*RefWorks:1040,RefWorks:1018}, on which we have to perform the so-called Schep correction~\cite{RefWorks:1078,*RefWorks:1077,RefWorks:876,RefWorks:1281}
\begin{equation}
\frac{1}{g^{\uparrow \downarrow}} = \frac{1}{\tilde{g}^{\uparrow \downarrow}}- \frac{1}{2g_{sh}}.
\label{eq:Schep_correction}
\end{equation}
In general the spin mixing conductance is a complex number, however, first principles calculations and experimental results on ferromagnetic resonance field shift show that the imaginary component is low for metallic interfaces~\cite{RefWorks:1046}, and therefore we mean only the real part here. According to Ref.~\citenum{RefWorks:1319}, for Pt, ${g}^{\uparrow \downarrow}$ with spin-orbit coupling is calculated \emph{with} Schep correction as $(h/e^2)1.07e15\,m^{-2}$ and $g_{sh}=(h/e^2)1e15\,m^{-2}$.

The effective spin mixing conductance of the spin circuit in Fig.~\ref{fig:spin_pumping_circuit}(b) $G_{eff} = G_{MIX} || (G_I + G_{SHE}) = lw \, (2e^2/h)\, g_{eff}^{\uparrow \downarrow}$, where
\begin{equation}
g_{eff}^{\uparrow \downarrow} = \frac{g^{\uparrow \downarrow}(g_I + g_{SHE})}{g^{\uparrow \downarrow} + (g_I + g_{SHE})}.
\label{eq:g_mix_eff}
\end{equation}
The above equation can be written as
\begin{equation}
g^{\uparrow \downarrow} = \frac{g_{eff}^{\uparrow \downarrow}}{1-g_{eff}^{\uparrow \downarrow}/(g_I + g_{SHE})}.
\label{eq:g_mix}
\end{equation}
Since $g^{\uparrow \downarrow} > 0$, we can write
\begin{equation}
g_{eff}^{\uparrow \downarrow} < g_I + g_{SHE},
\label{eq:geff_ineq}
\end{equation}
which is true since the the effective conductance $g_{eff}^{\uparrow \downarrow}$ of the circuit presented in Fig.~\ref{fig:spin_pumping_circuit}(b) cannot be grater than the conductances due to spin memory loss at the interface and of the bulk SHE layer. 

If the spin memory loss at the interface is negligible, i.e., $g_I \simeq 0$ ($\delta=0$, $sinh(\delta)=0$, and $cosh(\delta)=1$), given an experimentally obtained value of $g_{eff}^{\uparrow \downarrow}$ and other parameters like thickness $t$ and conductivity $\sigma$ of the SHE layer, we can have a maximum critical possible value $\lambda_{crit}= (h/2e^2)\, (\sigma/g_{eff}^{\uparrow \downarrow})$. From Ref.~\citenum{RefWorks:986}, $g_{eff}^{\uparrow \downarrow}=2.5e19\,m^{-2}$ at $t$ = 65 nm, and $\sigma=4.3e6$ 1/$\Omega$-m, which gives $\lambda_{crit}$ = 2.19 nm. In Refs.~\citenum{RefWorks:814,*RefWorks:813}, $g_{eff}^{\uparrow \downarrow}=2.1e19\,m^{-2}$ at $t$ = 15 nm, and $\sigma=2.4e6$ 1/$\Omega$-m, which gives $\lambda_{crit}$ = 1.475 nm, however, $\lambda$ is chosen as 10 nm therein, which is apparently inconsistent according to the underlying theoretical constructs as explained above~\cite{roy17_2,*roy_aps_2015}. It should be noted that with a significant spin memory loss (i.e., high $g_I$), the Equation~\eqref{eq:geff_ineq} does always tend to be satisfied.

From Equation~\eqref{eq:g_mix_eff}, note that $g_{eff}^{\uparrow \downarrow}$ is thickness-dependent due to the thickness dependence of $g_{SHE}$ and the trend depends on how $t/\lambda$ scales with lowering thickness, while $\lambda \propto \sigma$ according to Elliott-Yafet spin relaxation mechanism. With $t$ approaching zero $g_{eff}^{\uparrow \downarrow}$ must go down to zero, because both $g_{SHE}$ and $g_I$ (note $\delta=t_I/\lambda_I$, where $t_I$ and $\lambda_I$ are the interface thickness and interface spin diffusion length, respectively, goes toward zero too) go toward zero. It is possible to measure experimentally both the $g_{eff}^{\uparrow \downarrow}$ (from the enhancement of damping) and conductivity $\sigma$ with thickness $t$. Then choosing a value of $\lambda_{max}$ (at very high thickness $t >> \lambda$) and $(R^*,\delta)$ representing interfacial spin memory loss, the $g^{\uparrow \downarrow}$ can be calculated from the Equation~\eqref{eq:g_mix}. Using $(\lambda_{max}$, $g^{\uparrow \downarrow})$, $(R^*,\delta)$, and the relation $\lambda(t) \propto \sigma(t)$, we can calculate $g_{eff}^{\uparrow \downarrow}(t)$ from the Equation~\eqref{eq:g_mix_eff}. We can choose the $\lambda_{max}$ and $(R^*,\delta)$ that give us the best fit with the experimental data and the corresponding $g^{\uparrow \downarrow}$ to characterize the experimental results of $g_{eff}^{\uparrow \downarrow}(t)$.

Note that the total effective spin-mixing conductance $g_{eff}^{\uparrow \downarrow}$ (that includes any possible spin memory loss), which can be experimentally determined from the enhancement of damping due to spin pumping in ferromagnetic resonance experiments~\cite{RefWorks:881,RefWorks:1041,RefWorks:876}, is the one that is important for the measurement of the inverse spin Hall voltage $V_{ISHE}$ along the length of the SHE layer. Ref.~\citenum{RefWorks:986} defines a metric $\tilde{V}^{SP}_{ISHE}$ from the measurement of $V_{ISHE}$ and use the relation
\begin{equation}
\frac{\tilde{V}^{SP}_{ISHE}}{e R w g_{eff}^{\uparrow \downarrow}f} = \theta_{SH} \lambda \, tanh\left(\frac{t}{2\lambda}\right),
\label{eq:V_tilde_SP_ISHE}
\end{equation}
where $\theta_{SH}$ is the spin Hall angle, and $R$ is the resistance of the bilayer, which is inverse of $(\sigma t + \sigma_m t_m)\,w/l$ ($\sigma_m$ is the conductivity of the magnetic layer). Note that such expression is similar to the expression derived in Ref.~\citenum{RefWorks:814,*RefWorks:813}. The frequency dependent elliptical precession factor $P$ (Ref.~\citenum{RefWorks:884}) used in Ref.~\citenum{RefWorks:813} is included in the metric $\tilde{V}^{SP}_{ISHE}$ in  Ref.~\citenum{RefWorks:986} and attributed to the in-pane and out-of-plane precessing angles. Note that the metric defined by the Equation~\eqref{eq:V_tilde_SP_ISHE} is quite frequency-independent as the ferromagnetic resonance measurement results show in Ref.~\citenum{RefWorks:986} for 8 and 9 GHz. Also, note that the spin rectification voltage needs to be separated from the inverse spin Hall voltage $V_{ISHE}$~\cite{RefWorks:814,*RefWorks:813,RefWorks:986,RefWorks:984,*RefWorks:985}.

\section{Results and Discussions}

Figure~\ref{fig:match_experiments} shows the fitting of the experimental results $g_{eff}^{\uparrow \downarrow}$ and $\tilde{V}^{SP}_{ISHE}/eRwg_{eff}^{\uparrow \downarrow}f$ from Ref.~\citenum{RefWorks:986}. The Equations~\eqref{eq:g_mix_eff} and~\eqref{eq:V_tilde_SP_ISHE} have been used to match the experimental results. The thickness $t$ dependence of conductivity $\sigma$ and spin diffusion length $\lambda$, which is plotted in Fig.~\ref{fig:lambda_sigma}, is utilized here to successfully match the experimental results of $g_{eff}^{\uparrow \downarrow}$ and $\tilde{V}^{SP}_{ISHE}/eRwg_{eff}^{\uparrow \downarrow}f$ \emph{simultaneously}. Otherwise, according to Ref.~\citenum{RefWorks:986}, it would have taken two thickness-independent spin diffusion lengths ($\sim$1.5 nm and 8.3 nm) to match $g_{eff}^{\uparrow \downarrow}$ and $\tilde{V}^{SP}_{ISHE}/eRwg_{eff}^{\uparrow \downarrow}f$, respectively, which is unreasonable and termed as a controversy therein. We get from Ref.~\citenum{RefWorks:1319} $g^{\uparrow \downarrow} = 2.76e19\,m^{-2}$, and $g_I=1.11e19\,m^{-2}$ using $\delta=3.7$ and $R^*/\delta=23.6\,f\Omega m^2$ (i.e., $R^*=87.3\,f\Omega m^2$) at $\sigma$ = 4.3e6 1/$\Omega$-m. According to Ref.~\citenum{RefWorks:986}, $g_{eff}^{\uparrow \downarrow}=2.5e19\,m^{-2}$ at $t$ = 65 nm, and $\sigma$ = 4.3e6 1/$\Omega$-m. Hence we determine $\lambda_{max}$ = 3.67 nm at high thickness regime $t >> \lambda$ using the following equation.

\begin{equation}
\lambda_{max}=\cfrac{h}{2e^2}\, \cfrac{\sigma_{max} cosh(\delta)}{g^{\uparrow \downarrow} g_{eff}^{\uparrow \downarrow}/(g^{\uparrow \downarrow}-g_{eff}^{\uparrow \downarrow}) - g_I}.
\label{eq:lambda}
\end{equation}

Figure~\ref{fig:lambda_sigma} shows the thickness $t$ dependence of conductivity $\sigma$ and spin diffusion length $\lambda$ with $\lambda \propto \sigma$ signifying Elliot-Yafet spin relaxation mechanism. The Ref.~\citenum{RefWorks:986} specifies the experimental value of  conductivity 4.3e6 1/$\Omega$-m, which is a relevant saturated bulk value of conductivity~\cite{RefWorks:1111}. However, since the Ref.~\citenum{RefWorks:986} does not provide any thickness dependent data of $\sigma$, the trend of conductivity with thickness is taken from the experimental data provided in Ref.~\citenum{RefWorks:885} with a fitting of $\sigma = 4.3e6 \times (1-e^{-t/10.84 nm})$. Similar fitting model is used in Ref.~\citenum{RefWorks:1111}. As calculated earlier, the value of $\lambda_{max}$ ($\sigma_{max}$) = 3.67 nm (4.3e6 1/$\Omega$-m). 

Figure~\ref{fig:lambda_constant_compare} compares the results if we would have assumed the Dyakonov-Perel spin relaxation mechanism, with $\lambda = \lambda_{max} = 3.67$ nm. The comparison shows pretty clearly that two different spin diffusion lengths ($\sim$1.5 nm and 8.3 nm) would have been necessary to match the parts (a) and (b), respectively, according to Ref.~\citenum{RefWorks:986} alongwith the assumption that conductivity $\sigma$ is independent of thickness. Therefore, the assumptions are inconsistent in different ways and we do not assume that $\sigma$ is constant with $t$ in Fig.~\ref{fig:lambda_constant_compare}. As shown in the Fig.~\ref{fig:match_experiments}, $\lambda \propto \sigma$ (signifying Elliott-Yafet spin relaxation mechanism) gives the correct fit for both the parts (a) and (b), \emph{simultaneously}.

\begin{figure*}
\centering
\includegraphics{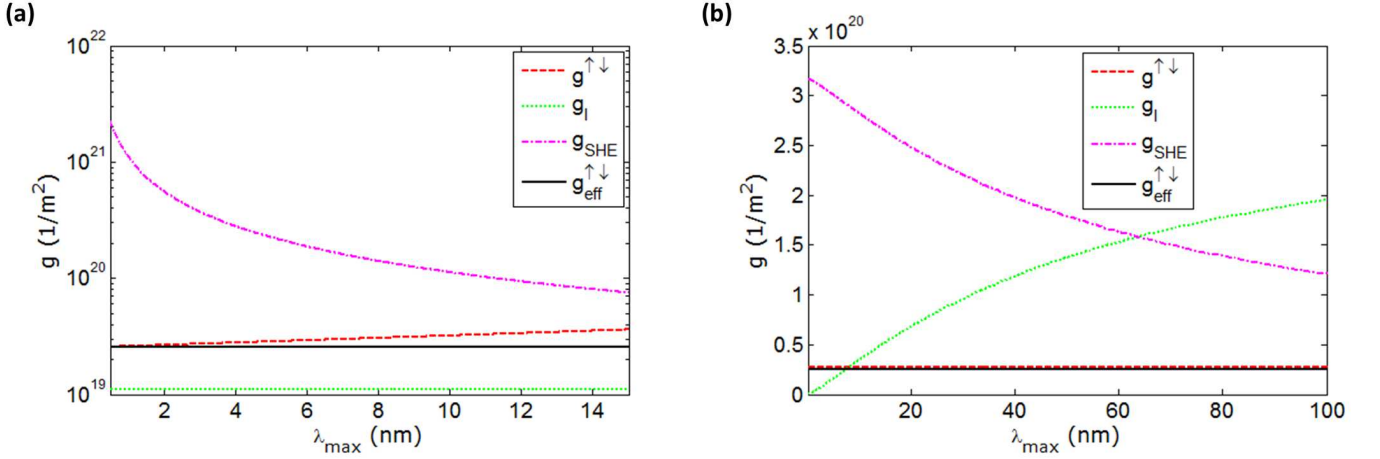}
\caption{\label{fig:g_lambda} (a) The $\lambda_{max}$ (0.5--15 nm) dependence of the $g^{\uparrow \downarrow}$ and $g_{SHE}$ for a constant effective spin mixing conductance $g_{eff}^{\uparrow \downarrow}$ and constant $g_I $ representing the interfacial spin memory loss. The $g^{\uparrow \downarrow}$ increases with increasing $\lambda_{max}$ while $g_{SHE}$ follows the opposite trend. (b) The $\lambda_{max}$ dependence of $g_I $ and $g_{SHE}$ for constant values of $g_{eff}^{\uparrow \downarrow}$ and $g^{\uparrow \downarrow}$. The spin flip parameter $\delta$ is varied from 0 to 6.07 while the interface resistance is kept constant at $R^*=87.3\,f\Omega m^2$. Hence, the $\lambda_{max}$ varies from 0.18 nm to 98 nm. The $g_I$ increases with increasing $\lambda_{max}$ while $g_{SHE}$ follows the opposite trend. The thickness $t$ of the SHE layer for both the cases is 65 nm.
}
\end{figure*}

\begin{figure}
\centering
\includegraphics{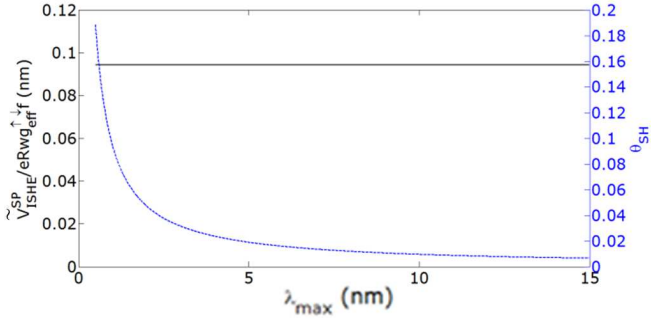}
\caption{\label{fig:SHA_lambda} The $\lambda_{max}$ (0.5--15 nm) dependence of the spin Hall angle $\theta_{SH}$ while keeping the metric $\tilde{V}^{SP}_{ISHE}/e R w g_{eff}^{\uparrow \downarrow}f = \theta_{SH} \lambda \, tanh(t/\lambda)$ [Equation~\eqref{eq:V_tilde_SP_ISHE}]  constant. The thickness $t$ is assumed to be 65 nm. The $\theta_{SH}$ decreases with increasing $\lambda_{max}$ keeping $\theta_{SH} \lambda_{max}$ constant.}
\end{figure}

Figure~\ref{fig:g_lambda}(a) shows the $\lambda_{max}$ (0.5--15 nm) dependence of the different conductances involved while keeping $g_{eff}^{\uparrow \downarrow}$ constant, since this is directly obtained from experiments. For a certain value of $g_{eff}^{\uparrow \downarrow}$, as $\lambda_{max}$ increases, the spin conductance of the SHE layer $g_{SHE}$ decreases, and therefore the interfacial spin mixing conductance $g^{\uparrow \downarrow}$ increases. (The conductance due to interfacial spin memory loss $g_I$ is kept constant.) However, the change in $g^{\uparrow \downarrow}$ is less than 3\% for the range of $\lambda_{max}$ with reference to the nominal value of $g^{\uparrow \downarrow} = 2.76e19\,m^{-2}$ at $\lambda_{max}$ = 3.67 nm. Therefore, a slight variation in $g^{\uparrow \downarrow}$ can lead to a large variation in $\lambda_{max}$. 

Figure~\ref{fig:g_lambda}(b) shows the trend of $g_I$ and $g_{SHE}$ with $\lambda_{max}$ when the interface spin flip parameter $\delta$ varied between 0 and 6.07, and $R^*$, $g_{eff}^{\uparrow \downarrow}$, and $g^{\uparrow \downarrow}$ are kept constant. The range of $\lambda_{max}$ turns out to be 0.18 nm ($g_I$ = 0) -- 98 nm ($g_I=1.94e20\,m^{-2}$). The $g_{SHE}$ decreases (with the increase of both $\delta$ and $\lambda_{max}$) with increasing $\lambda_{max}$ so that the sum $g_I+g_{SHE}$ is constant at $3.168e20\,m^{-2}$. With sufficient increase of $\delta$, $g_I$ starts to take over $g_{SHE}$ at $\delta=5.9$ and $\lambda_{max}$ = 64.32 nm. A further increase of $\delta$ makes $g_I$ equal to $3.168e20\,m^{-2}$, the sum of $g_I+g_{SHE}$ with $g_{SHE}$ tending to zero, and such maximum value $\delta_{max}$ can be calculated as 6.4919 by solving the following equation.

\begin{equation}
\delta_{max}sinh(\delta_{max})=\cfrac{2e^2}{h}\, R^* \, \cfrac{g^{\uparrow \downarrow} g_{eff}^{\uparrow \downarrow}}{g^{\uparrow \downarrow}-g_{eff}^{\uparrow \downarrow}}.
\label{eq:delta}
\end{equation}

Figure~\ref{fig:SHA_lambda} shows that the spin Hall angle $\theta_{SH}$ decreases inversely proportional to $\lambda_{max}$ for a certain value of the metric on the left-hand side of the Equation~\eqref{eq:V_tilde_SP_ISHE}. Therefore, for a given $\lambda_{max}$, we get $\theta_{SH}$. Hence, we can get the set $(\lambda_{max},\theta_{SH},g^{\uparrow \downarrow})$ as (3.67 nm, 0.026, 2.76e19 $m^{-2}$) that we have used to match the experimental results as depicted in the Fig.~\ref{fig:match_experiments}. However, these two sets (7 nm, 0.014, 2.98e19 $m^{-2}$) and (0.50 nm, 0.19, 2.57e19 $m^{-2}$) would also match the experimental results in the Fig.~\ref{fig:match_experiments}. It should be noted that the factor $t/\lambda(t)$ at $t=2$ nm for the aforesaid three sets are 3.04 [$\lambda$ (2 nm) = 0.66 nm], 1.69 [$\lambda$ (2 nm) = 1.18 nm], and 23.68 [$\lambda$ (2 nm) = 0.08 nm], respectively, making the $tanh(t/\lambda)$ term greater than 0.9. For all these calculations $g_I=1.95e18\,m^{-2}$. Therefore an understanding on any one parameter of the set $(\lambda_{max},\theta_{SH},g^{\uparrow \downarrow})$ is required to get the complete set. We have utilized $g^{\uparrow \downarrow}$ from first-principles calculations~\cite{RefWorks:1319} and earlier described in the Section~\ref{sc:model} that experimental data on thickness-dependent $g_{eff}^{\uparrow \downarrow}$ (at low-thickness regime where $g_{eff}^{\uparrow \downarrow}$ varies) can provide us an accurate set $(\lambda_{max},g^{\uparrow \downarrow},g_I)$. Also, it may be possible to know $\theta_{SH}$ from the spin-torque ferromagnetic resonance~\cite{RefWorks:817,RefWorks:1016,RefWorks:1333}, or $\lambda$ from other experiments~\cite{RefWorks:1316,RefWorks:1013,RefWorks:1322}.

\section{Summary}

To summarize, we have shown that a thickness-dependent spin diffusion length for platinum signifying Elliott-Yafet spin relaxation mechanism is necessary to \emph{simultaneously} match the experimental results of thickness-dependent effective spin mixing conductance and the inverse spin Hall voltage induced by spin pumping. Similar analysis can be applied to other SHE materials e.g., palladium (Pd)~\cite{RefWorks:1286,RefWorks:813,RefWorks:982,RefWorks:1299,RefWorks:811}, tantalum (Ta)~\cite{RefWorks:608,RefWorks:1021,RefWorks:816,RefWorks:811,RefWorks:1344,RefWorks:780}, and tungsten (W)~\cite{RefWorks:758}. We note that the point of having significant interfacial spin memory loss (a significant loss makes the spin diffusion length higher) has controversy in literature~\cite{RefWorks:367,RefWorks:1042,RefWorks:1340,RefWorks:1316,RefWorks:1114,RefWorks:1319,RefWorks:1113,RefWorks:1016,RefWorks:1351}, however, as analyzed, that does not change the conclusion presented in this paper. The sample quality, i.e., conductivity can apparently result in large variation in spin diffusion length and spin Hall angle~\cite{RefWorks:1347}. Note that the spin Hall angle meant here is an effective one since interface spin Hall effect can be different from the bulk counterpart in general~\cite{RefWorks:1349,RefWorks:1348}. It needs to also carefully consider the low-thickness regime ($<$ 2 nm), due to magnetic proximity effect~\cite{RefWorks:1005,*RefWorks:1006,RefWorks:1312,RefWorks:1007,RefWorks:988,RefWorks:1323}. Since the estimated parameters are sensitive to the variation in interface conductances, variation tolerant design principles may need to be employed for engineering applications. The comprehensive analysis performed here has immense consequence on device design and predicting correct device functionality for potential technological applications.

\section*{Acknowledgements}
%%\noindent {\bf Acknowledgements}: 
%\vspace*{2mm}
This work was supported by FAME, one of six centers of STARnet, a Semiconductor Research Corporation program sponsored by MARCO and DARPA. 

%\bibliographystyle{aipnum4-1}
%%\nocite{*}
%\bibliography{royk,royk2}% Produces the bibliography via BibTeX.
%merlin.mbs aipnum4-1.bst 2010-07-25 4.21a (PWD, AO, DPC) hacked
%Control: key (0)
%Control: author (8) initials jnrlst
%Control: editor formatted (1) identically to author
%Control: production of article title (-1) disabled
%Control: page (0) single
%Control: year (1) truncated
%Control: production of eprint (0) enabled
%

\end{document}